\documentclass[aps,prl,article,showpacs,preprintnumbers,twocolumn,amsmath,amssymb,superscriptaddress]{revtex4-1}
\date{\today}
\usepackage{epsfig}
\usepackage{subfigure}
\usepackage{graphicx}
\usepackage{dcolumn}
\usepackage{bm}
\usepackage[colorlinks,linkcolor=blue,hyperindex,CJKbookmarks]{hyperref}
\usepackage{float}
\usepackage{hyperref}
\usepackage{comment}
\usepackage{color}
\usepackage{ulem}
\begin{document}

\def\rc{$\alpha$-RuCl$_{3}$}

\title{Low temperature enhancement of ferromagnetic Kitaev correlations in \rc}

\author{Andreas Koitzsch}
\affiliation{IFW Dresden, Helmholtzstra\ss e 20, 01069 Dresden, Germany}
 
\author{Eric M\"{u}ller}
\affiliation{IFW Dresden, Helmholtzstra\ss e 20, 01069 Dresden, Germany}
 
\author{Martin Knupfer}
\affiliation{IFW Dresden, Helmholtzstra\ss e 20, 01069 Dresden, Germany}

\author{Bernd B\"{u}chner}
\affiliation{IFW Dresden, Helmholtzstra\ss e 20, 01069 Dresden, Germany}
\affiliation{Department of Physics, TU Dresden, 01069 Dresden, Germany}

\author{Domenic Nowak}
\affiliation{Department of Chemistry and Food Chemistry, TU Dresden, 01069 Dresden, Germany}

\author{Anna Isaeva}
\affiliation{Department of Chemistry and Food Chemistry, TU Dresden, 01069 Dresden, Germany}

\author{Thomas Doert}
\affiliation{Department of Chemistry and Food Chemistry, TU Dresden, 01069 Dresden, Germany}

\author{Markus Gr\"{u}ninger}
\affiliation{II. Physikalisches Institut, Universit\"{a}t zu K\"{o}ln, Z\"{u}lpicher Strasse 77, D-50937 K\"{o}ln, Germany}

\author{Satoshi Nishimoto}
\affiliation{IFW Dresden, Helmholtzstra\ss e 20, 01069 Dresden, Germany}
\affiliation{Department of Physics, TU Dresden, 01069 Dresden, Germany}

\author{Jeroen van den Brink}
\affiliation{IFW Dresden, Helmholtzstra\ss e 20, 01069 Dresden, Germany}
\affiliation{Department of Physics, TU Dresden, 01069 Dresden, Germany}

\date{\today}
\begin{abstract}
Kitaev-type  interactions between neighbouring magnetic moments emerge in the honeycomb material \rc. 
It is debated however whether these Kitaev interactions are ferromagnetic or antiferromagnetic. 
With electron energy loss spectroscopy (EELS) we study the lowest excitation across the Mott-Hubbard gap, which involves a $d^4$ triplet in the final state and therefore is sensitive to nearest-neighbor spin-spin correlations. 
At low temperature the spectral weight of these triplets is strongly enhanced, in accordance with optical data. 
We show that the magnetic correlation function that determines this EELS spectral weight is directly related to a Kitaev-type spin-spin correlator and that 
the temperature dependence agrees very well with the results of a microscopic magnetic Hamiltonian for \rc\ with {\it ferromagnetic} Kitaev coupling.
\end{abstract}
\maketitle

{\it Introduction ---}
The celebrated Kitaev model describes bond-dependent spin 1/2 interactions on the honeycomb-lattice \cite{Kitaev2006}.
It has attracted enormous attention because it is conceptually simple but harbors rich physics and is still exactly solvable. Among its solutions are quantum spin liquids which show a number of peculiar properties, such as the absence of magnetic long range order at $T=0$ despite the presence of sizable moments and exotic fractionalized excitations like Majorana fermions with potential applications for quantum information processing. 

After the identification of iridates as possible solid state realizations of the Kitaev model \cite{Jackeli2009} much work has been devoted to Ir$^{4+}$ systems, with its 5$d^5$ electron configuration and the effective $J_{eff}=1/2$ description in order to uncover signatures of the quantum spin liquid \cite{Kim2008, Rau2014}. However, research on the iridates is hampered by e.g. the difficult crystal growth and lattice distortions. Recently, \rc\ has been established as a promising $4d$ analogue to the iridates \cite{Plumb2014, Banerjee2016}. Neutron and Raman scattering studies gave evidence for fractionalized excitations typical for the Kitaev quantum spin liquid \cite{Banerjee2016,Sandilands2015b,Banerjee2017a,Banerjee2017b,Nasu2017} and both very recent theoretical \cite{Yadav2016a} and experimental investigations \cite{Baek2017,Wolter2017,Hentrich2017,Zheng2017,Leahy2016,Hentrich2017,Wang2017,Jansa2017,Ponomaryov2017,Banerjee2017b} indicate in this material the presence of a transition into a quantum spin liquid state in an external magnetic field.

Thus the quantification of the bond-dependent Kitaev interaction term $K$ has become a key issue for \rc. For a deeper understanding and correct theoretical description of the material properties knowledge of $K$ is crucial, much like knowledge of the Heisenberg exchange parameter $J$ for ordinary magnets. Unfortunately, in spite of extensive and detailed investigations of its electronic and magnetic structure \cite{Kubota2015,Majumder2015,Sears2015,Kim2015a,Rousochatzakis2015,Zhou2016b,Koitzsch2016a,Kim2016,Cao2016,Winter2016,Lang2016,Janssen2016,Sizyuk2016,Ziatdinov2016,Sinn2016,Sandilands2016,Sandilands2016b,Banerjee2017a,Banerjee2017b,Nasu2017,Ran2017,Agrestini2017,Hou2017,Winter2017} 
presently not even the sign of $K$ is known with certainty for \rc. Whereas Banerjee et al. (Ref.~\onlinecite{Banerjee2016,Banerjee2017a}) introduce positive $K$, that is antiferromagnetic coupling, to fit spin-wave spectra measured by inelastic neutron scattering, quantum chemistry studies favor negative (ferromagnetic) $K$ \cite{Yadav2016a}. Other neutron scattering experiments indeed claim better agreement with a ferromagnetic $K$ \cite{Ran2017,Banerjee2017b}.

Here we shed light on this controversy by measuring the temperature dependent loss function by electron energy loss spectroscopy (EELS) and comparing the results to the spectral weight derived from a microscopic Hamiltonian. We conclude from this that as the temperature goes down the nearest neighbor Kitaev spin-spin correlations become more ferromagnetic pointing unequivocally to a ferromagnetic $K$ in the Hamiltonian for \rc.   

\begin{figure}
\includegraphics[width=.9\linewidth]{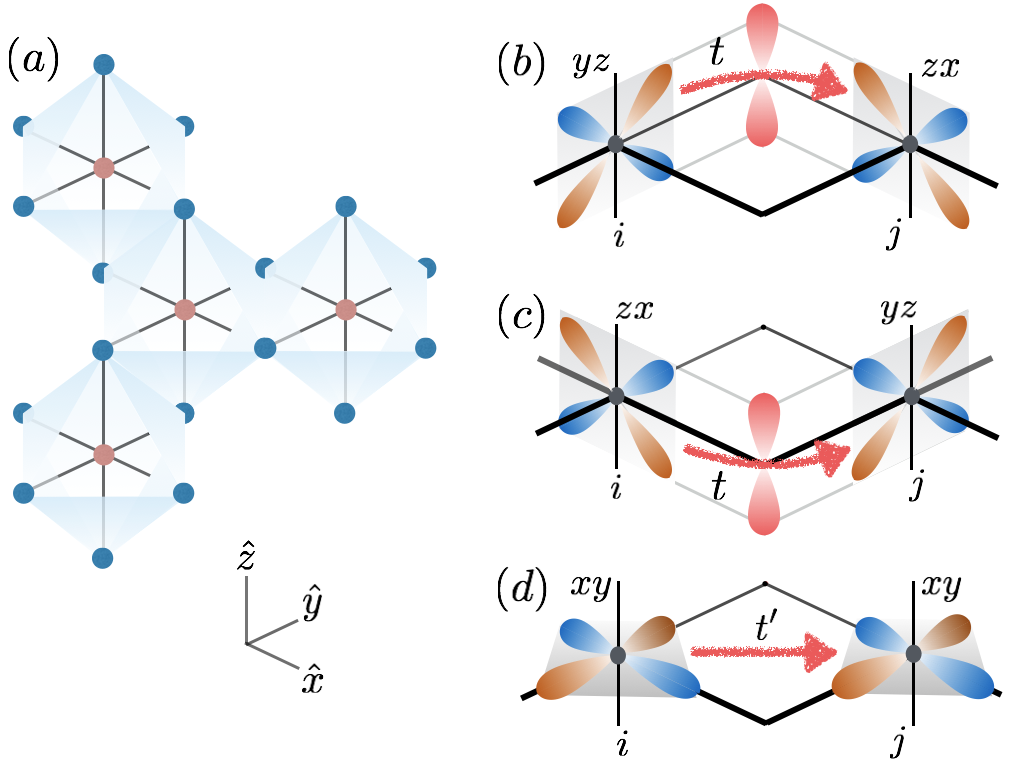}
\caption{(a) Illustration of edge-sharing [RuCl$_6$] octahedra comprising a single honeycomb layer of $\alpha$-RuCl$_3$ and representation of the hopping integrals between Ru $t_{2g}$ orbitals of $d_{yz}$, $d_{zx}$ and $d_{xy}$ symmetry. Note that along the two different paths with amplitude $t$ from $i$ to $j$ indicated in (b) and (c) different orbitals are involved; (d) direct hopping $t'$ between $d_{xy}$ states. 
}
\label{fig1}
\end{figure}

{\it Conceptual background ---}
In $\alpha$-RuCl$_3$ the basic structural building blocks, [RuCl$_6$] octahedra, are connected via edges into layers propagating in the $ab$ plane, see Fig.~\ref{fig1}a. The $d^5$ configuration of Ru$^{3+}$ has a single hole in the $t_{2g}$ shell, where the wavefunctions of the three $t_{2g}$ orbitals $d_{xy}$, $d_{yz}$ and $d_{zx}$ are indicated in Fig.~\ref{fig1}.  The strong spin-orbit coupling splits the $t_{2g}$ states into a $j=3/2$ quartet and $j=1/2$ Kramers doublet, the latter forming the ground state.
It is important to note that for edge-sharing octahedra the hopping between the $t_{2g}$ orbitals via the ligands has a very specific  symmetry \cite{Jackeli2009}:
the largest hopping $t$ is between the $d_{yz}$ and $d_{zx}$ orbitals on neigboring sites, as indicated in Fig.~\ref{fig1}b-c. 
Additional hopping amplitudes are symmetry allowed, but tend to be much weaker -- an example is the direct hopping  $t'$  between  $d_{xy}$ orbitals on neighboring sites (see  Fig.~\ref{fig1}d).

Orbital-dependent superexchange interactions~\cite{Kugel1984,Khaliullin2005} of these spin-orbit coupled Ru$^{3+}$ magnetic moments are generated by inter-site hopping processes of the type $d^5$-$d^5$ $\rightarrow$ $d^4$-$d^6$ $\rightarrow$ $d^5$-$d^5$. It turns out that in leading order ($t^2/U$, with hopping $t$ and Hubbard $U$) the exchange interactions vanish, but  in next order  ($t^2J_H/U^2$, with Hund's rule coupling $J_H$) the interactions are precisely of the bond-directional type as they appear in the Kitaev Hamiltonian~\cite{Jackeli2009,Chaloupka2010}. The essential ingredient that causes the Kitaev coupling $K$ to become finite is the fact that $J_H$ splits up the $d^4$ intermediate states, for which there are two holes on the same site, into a local manifold of triplets ($^3T_1$, 9 states) and singlets ($^1T_2$, $^1E$, $^1A$, 6 states). As $J_H$ is ferromagnetic, the $^3T_1$ multiplet is well below the singlet states in energy.

These hopping amplitudes not only determine the form and magnitude of the exchange interactions, they also determine the spectral weight of intersite $d^5$-$d^5$ $\rightarrow$ $d^4$-$d^6$  excitations \cite{Khaliullin2005,Chaloupka2010,Kovaleva2004,Lee2005,Goessling2008,Reul2012}, i.e., excitations across the Mott-Hubbard gap as measured in for instance EELS and optical spectroscopy. For the lowest $d^4$-$d^6$ excited states, it is sufficient to consider a single $d^6$ ($t_{2g}^6$) multiplet and the lowest $d^4$ multiplets. Accordingly, the intersite excitations in RuCl$_3$ are also split into the $d^4$ triplet and singlet multiplets, i.e., they are directly related to the intermediate states of the superexchange interaction, and their spectral weight is linked in particular to magnetic correlations of Kitaev-type, as we will quantify in the following.

{\it Electron energy loss measurements ---}
Our EELS experiments on RuCl$_3$ were performed on platelet-like single crystals up to several mm in diameter, grown by chemical vapor transport reactions, see Ref.~\onlinecite{Koitzsch2016a} for details of crystal growth and characterization. The EELS measurements were carried out using a purpose built transmission electron energy-loss spectrometer \cite{Fink1989, Roth2014} with a primary electron energy of 172 keV and energy and momentum resolutions of $\Delta$\textit{E} = 85 meV and $\Delta$\textbf{q} = 0.035 \AA$^{-1}$, respectively. The films ($\textit{d} \approx$ 100 nm) were exfoliated by scotch tape. Subsequently, the films were mounted onto standard electron microscopy grids and transferred into the EELS spectrometer.

\begin{figure}[t]
\includegraphics[width=1\linewidth]{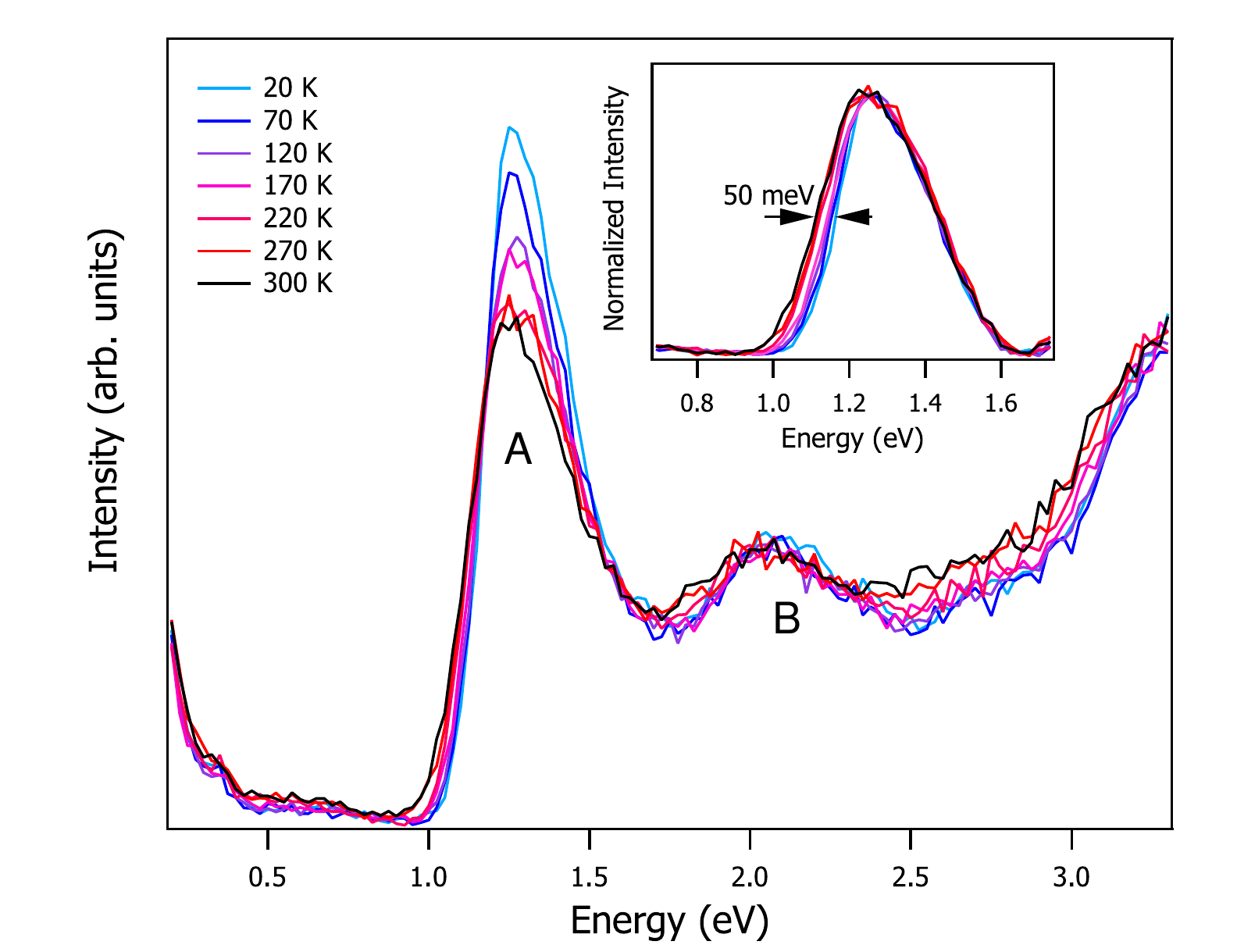}%
\caption{Temperature dependent low-energy loss function measured at $\textbf{q}=0.1$ \AA$^{-1}$. Inset: peak A background corrected and normalized.}
\label{fig2}
\end{figure}

\begin{figure}
\includegraphics[width=\columnwidth]{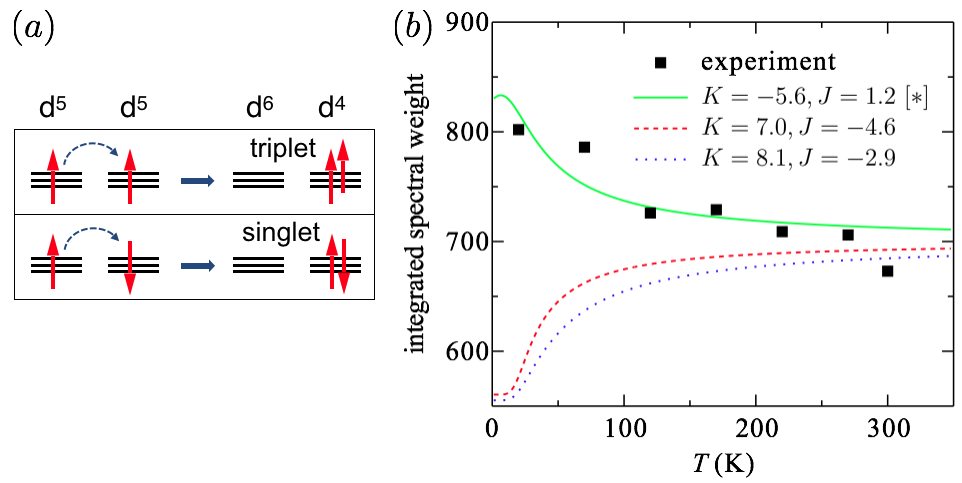}
\caption{ (a) Schematics of the Mott excitation involving the $t_{2g}$ shells of two neighboring Ru$^{3+}$ sites. Red arrows refer to holes.  (b)  Integrated spectral weight $I$ of peak A in Fig. 2 after background subtraction as a function of temperature compared to the results of exact diagonalization calculations for different parametrizations of the magnetic interactions -- $[*]$ refers to the exchange parameters determined in Ref.~\onlinecite{Yadav2016a} (see also main text).
}
\label{fig3}%
\end{figure}

Figure~\ref{fig2} shows the low-energy loss function measured between $T=20$ K and 300 K in the quasi optical limit of small momentum transfer ($\textbf{q}=0.1$ \AA$^{-1}$). 
The spectra are normalized at higher energy ($E = 4$ eV).  The lowest EELS features can be assigned to optical excitations of the $d^5$-$d^5\rightarrow d^4$-$d^6$ type across the Mott-Hubbard gap of 1.1 eV. We observe peaks at $E_A=1.2$ eV and  $E_B=2.1$ eV, consistent with our previous studies \cite{Koitzsch2016a} and with optical conductivity data \cite{Sandilands2016}. 
As indicated in Fig.~\ref{fig3}a the EELS $d^4$-$d^6$ final state may contain a $d^4$ spin singlet ($S=0$) or triplet ($S=1$) state, which significantly differ in energy due to Hund's rule, i.e., interorbital exchange interaction $J_H$. The triplet is expected at much lower energy, $2J_H \approx 0.8$ eV, thus the lowest-energy feature, peak A, is associated with the $d^4$ triplet multiplet, in agreement with previous reports \cite{Sandilands2016b,Koitzsch2016a}. Peak B has been assigned to either the $d^4$ singlet multiplet \cite{Sandilands2016b} or to a crystal-field excitation 
to empty $e_g$ states \cite{Koitzsch2016a}. Our analysis is focused on peak A because a quantitative correlation between the spectral weight and the nearest-neighbor spin-spin correlations requires a clear separation between different multiplets, which typically is only realized for the lowest excitation across the gap. This is supported by the temperature dependence. The spectral weight of peak A decreases significantly with increasing temperature, see Fig.~\ref{fig3}b, in agreement with recent optical conductivity data \cite{Sandilands2016b}. In contrast, peak B does not show such a clear temperature dependence.  

{\it Qualitative interpretation ---}
It is well known that nearest-neighbors spin-spin correlations in Mott-Hubbard insulators may cause large spectral weight changes across magnetic phase transitions at $T_N$ even when $k_BT_N$ is much smaller than the energy gap \cite{Khaliullin2004,Kovaleva2004,Lee2005,Reul2012}. This simply reflects the spin selection rule for optical excitations. As will be quantified below, excitations to $d^4$ triplets acquire finite spectral weight {\it only} if the initial alignment of the magnetic moments on two neighboring Ru sites is parallel, as is the case for a dominating ferromagnetic Kitaev exchange. The increase of spectral weight at lower temperature implies that the nearest-neighbor Kitaev spin-spin correlation becomes more ferromagnetic. This directly points to the Kitaev term in the Hamiltonian being ferromagnetic, in line with quantum chemistry calculations~\cite{Yadav2016a}. If one assumes on the other hand the scenario of dominant antiferromagnetic nearest-neighbor Kitaev exchange, the spectral weight of peak A should be suppressed at low temperature, which is at odds with the experimental data. 

According to the arguments above, upon increasing temperature the ferromagnetic Kitaev correlations should be reduced, giving rise to an increase of spectral weight of the spin singlet states at higher energy. Indeed, Fig.~\ref{fig2} does show spectral weight transfer from peak A to a broad region between $E \approx 1.5 - 3.5$ eV. 

The inset of Fig.~\ref{fig2} presents peak A in a normalized and background-subtracted fashion in order to monitor the temperature dependence of the line shape and peak position. The line shape is slightly asymmetric but remains almost constant with temperature except for a modest broadening. The broadening and gap change are of the order of 50 meV  and can be explained by conventional thermal effects. 
Figure \ref{fig2} clearly shows that the temperature-induced change of spectral weight is not caused by a change of the line shape of peak A. Therefore we may rule out an excitonic effect as origin of the spectral weight change, confirming our interpretation in terms of spin-spin correlations. 

{\it Quantitative relation between triplet weight and magnetic correlations ---}
We now wish to determine on a quantitative, microscopic basis how the spectral weight of the lowest energy triplet excitations depends on the relative orientation of neighboring $j=1/2$ moments. Following Ref.~\onlinecite{Jackeli2009}, we define creation operators for a hole in the $d_{xy}$, $d_{yz}$ and $d_{zx}$ orbital on site $j$ as $x^\dagger_{j \sigma}$, $ y^\dagger_{j \sigma}$ and $z^\dagger_{j \sigma}$ with spin $\sigma= \uparrow$ or $\downarrow$. The creation operator $a^\dagger_{\bar{\sigma}}$ for a $j=1/2$ doublet state with pseudospin $\bar{\sigma} = \bar{\uparrow}$ or $\bar{\downarrow}$ is 
\begin{align}
\begin{array}{ll}
a^\dagger_{\bar{\uparrow}} =  x^\dagger_{\uparrow}\sin \theta+ \cos \theta \left( i z^\dagger_{\downarrow} + y^\dagger_{\downarrow} \right)/{\sqrt 2}\\
a^\dagger_{\bar{\downarrow}} = x^\dagger_{\downarrow} \sin \theta  +  \cos \theta \left( i z^\dagger_{\uparrow} - y^\dagger_{\uparrow} \right)/{\sqrt 2}
\end{array}
\end{align}
The strong spin-orbit coupling puts the $j=3/2$ quartet at much higher energy. Note that $\tan \theta = 1/\sqrt{2}$ for the high-symmetry "cubic" $j=1/2$ states which have equal contributions of the three $t_{2g}$ orbitals. The corresponding hopping Hamiltonian on a $z$ bond
is
\begin{align}
H^0_{\langle ij \rangle}=  \sum_\sigma 
\left[  \left(t y^\dagger_{j \sigma} z_{i \sigma} +t z^\dagger_{j \sigma} y_{i \sigma} 
+ t' x^\dagger_{j \sigma} x_{i \sigma} \right) + h.c. \right]
\label{eq:H0}
\end{align}
The optical/EELS spectral function is generated by the response of the system to the current operator $H'$, which is obtained from Eq.~\ref{eq:H0} by substituting $(t,t') \rightarrow (it,it')$. The matrix elements that we wish to evaluate are of the type $\langle \psi_T |H' | {\bar \sigma_i \bar \sigma'_j} \rangle$, where $\psi_T$ are the spin triplet $d^4$ states. It is easy to show that the part of $H'$ that is governed by $t'$, the direct $d_{xy}$-$d_{xy}$ channel, only causes singlet $d^4$ excitations and is therefore irrelevant for triplet spectral weight.
A detailed calculation 
provides, along a $z$-bond,  
$
\sum_T|\langle \psi_T |H'/t | {\bar \downarrow \bar \downarrow} \rangle|^2 = \sum_T |\langle \psi_T |H'/t | {\bar \uparrow \bar \uparrow} \rangle|^2=  \cos^4 \theta +\frac{1}{2} \sin^2 \theta \cos^2 \theta 
$
and
$
\sum_T|\langle \psi_T |H'/t | {\bar \uparrow \bar \downarrow} \rangle|^2 = \sum_T |\langle \psi_T |H'/t | {\bar \downarrow \bar \uparrow} \rangle|^2=   \sin^2 \theta \cos^2 \theta.
$
Collecting terms, the total intensity $I_T$ of the triplets is apart from a constant term
\begin{align}
I_T=\sum_{T, \langle ij \rangle} |\langle \psi_T |H' | \bar\sigma_i \bar\sigma_j \rangle|^2 =  \frac{2t^2}{3}  \sum_{ \langle ij \rangle} (S^z_i S^z_j +1/4)
\label{eq:correlator_pure}
\end{align}
for "cubic" $j=1/2$ states ($\tan^2 \theta = 1/2$ and $\cos^2\theta = 2/3$)  
where $S^z_i S^z_j$ is the Kitaev term on the $z$-bond -- the other bonds follow by replacing $S^z$ by $S^y$/$S^x$ respectively. From this expression it is clear that the triplet spectral weight is maximum when the pseudospins are oriented ferromagnetically and, vice versa, smallest when neighboring $j=1/2$ moments are oriented antiparallel. This is in line with the general expectation that two parallel neighboring moments are more likely to be excited into a triplet state than a pair of antiparallel moments. 

{\it Evaluation of magnetic correlations ---}
To compare with the experimental data, we evaluated the temperature dependent correlator in Eq.~\ref{eq:correlator_pure} numerically for the extended Heisenberg-Kitaev Hamiltonian for RuCl$_3$ with all symmetry allowed nearest-neighbor couplings that on the 
$z$ bonds takes the form
\begin{align}
H^M_{\langle ij \rangle}= J {\bf S}_i \cdot {\bf S}_j + K S^z_i S^z_j +\sum_{\alpha \neq \beta} \Gamma_{\alpha \beta} (S^\alpha_i S^\beta_j + S^\beta_i S^\alpha_j), \nonumber
\label{eq:HM}
\end{align}
with appropriate permutations for the $x$ and 
$y$ bonds on the honeycomb lattice. We use the magnetic couplings derived from quantum chemistry calculations: 
$J=1.2$ meV, $K=-5.6$ meV, $\Gamma_{xy}=-1.2$ meV, $\Gamma_{zx}=-\Gamma_{xy}=-0.7$ meV, and further neighbor exchange $J_2=J_3=0.26$ meV~\cite{Yadav2016a}. Note that  $|K/J| = 4.6$ and $K$ is ferromagnetic. Full exact diagonalization calculations with a 16-site periodic cluster were performed -- due to the dominant Kitaev term, finite-size effects are small. At zero temperature the nearest-neighbor spin-spin correlation in our model is 0.1087 and it is close to 0.1323 in the ferromagnetic Kitaev limit (for Heisenberg $J$=0), where the finite-size effect disappears (also see Ref.~\onlinecite{Yadav2016a}). We calculated the expectation value of the correlator in the canonical ensemble; here, only excitations with momentum transfer ${\bf q}={\bf 0}$ that are relevant for EELS are summed up. The expectation value of the nearest-neighbor spin-spin correlation is obtained as an averaged one of the three bonds, i.e.,  $\langle S^x_i S^x_j \rangle$ for the $x$-bond, $\langle S^y_i S^y_j \rangle$ for the $y$-bond, and $\langle S^z_i S^z_j \rangle$ for the $z$-bond. More detailed information is given in Supplementary Material.

{\it Discussion ---} A direct comparison of the numerically evaluated spin-spin correlator in Eq.~\ref{eq:correlator_pure} with the temperature dependence of the EELS spectral weight of peak A is provided in Fig.~\ref{fig3}b. We obtain good agreement using magnetic interactions in the Hamiltonian previously derived on the basis of quantum chemistry calculations, in particular a ferromagnetic
Kitaev exchange $K=-5.6$ meV (see Ref.~\onlinecite{Yadav2016a}). The same calculations done with antiferro Kitaev exchange, using in particular the values of $K$ and $J$ suggested from neutron scattering in Ref.~\onlinecite{Banerjee2016}, provide temperature trends that clearly do not agree with the EELS data, as shown in Fig.~\ref{fig3}b. The gradual decrease of the spectral weight up to temperatures far above $T_N$ is typical for magnetic systems with enhanced quantum fluctuations such as two-dimensional systems \cite{Goessling2008a} and thus expected for the strongly frustrated Kitaev model. Actually the temperature scale that governs the reduction of the triplet spectral weight should roughly correspond to the energy scale of $|K|$. Based on solely the experimental data (see Supplementary Material) one obtains for this $|K| \approx 90$ K or 7.7 meV, which indeed is qualitatively in agreement with the detailed theory. 

{\it Conclusions ---}
The importance of the Kitaev exchange in \rc\ has been established on the basis of various of its magnetic properties -- the fractionalized excitations seen in inelastic neutron scattering \cite{Banerjee2016}, Raman spectroscopy \cite{Nasu2016}, consequences for the static magnetic order \cite{Kim2015a, Sizyuk2016, Winter2016} and the magnetic field induced transitions into a quantum liquid state \cite{Yadav2016a,Baek2017,Wolter2017,Nasu2017,Hentrich2017,Zheng2017,Leahy2016}. These approaches have so far not resolved the question whether the Kitaev exchange in \rc\ is actually ferromagnetic or antiferromagnetic.
Here we follow a different ansatz by analyzing how the spin-spin correlations affect the \textit{electronic} excitation spectrum. 
We show that the magnetic correlation function that determines EELS and optical spectral weight is directly related to a Kitaev-type spin-spin correlator. The experimental observation of triplet spectral weight increasing at low temperature implies {\it ferromagnetic} Kitaev-type correlations becoming stronger.
The measured temperature dependence of the EELS spectral weight agrees with calculations for a microscopic magnetic Hamiltonian for \rc\ with ferromagnetic Kitaev coupling. Calculations for systems with antiferromagnetic Kitaev coupling exhibit a temperature dependence that is opposite to the one experimentally observed. 

\begin{acknowledgments}
	Financial support by the 
	Deutsche Forschungsgemeinschaft (DFG) through the Collaborative Research Centers SFB 1143 and SFB 1238 
	is gratefully acknowledged.
\end{acknowledgments}


%

\end{document}